\documentclass[aps,twocolumn,amsmath,letterpaper]{revtex4-1}
\usepackage{amsmath}
\usepackage{ifpdf}
\usepackage{graphicx,natbib,amssymb}
\usepackage{amssymb}
\usepackage{graphicx}
\usepackage{array}
\usepackage{hhline}
\usepackage{longtable}
\usepackage[plainpages=false]{hyperref}
\usepackage{float}
\usepackage{url}
\newcommand{\ket}[1]{\left| #1 \right>} 
\newcommand{\bra}[1]{\left< #1 \right|} 

\begin{document}

    \title{Simulating cw-ESR Spectrum Using Discrete Markov Model of \\ Single Brownian Trajectory}

 \author{Efe Ilker}
 \address{Faculty of Engineering and Natural Sciences, Sabanc\i~University, Tuzla 34956, Istanbul, Turkey,}

\begin{abstract}
Dynamic trajectories can be modeled with a Markov State Model (MSM). The reduction of continuous space coordinates to discretized coordinates can be done by statistical binning process. In addition to that, the transition probabilities can be determined by recording each event in the dynamic trajectory. This framework is put to a test by the electron spin resonance (ESR) spectroscopy of nitroxide spin label in X- and Q- bands. Calculated derivative spectra from MSM model with transition matrix obtained from a single Brownian trajectory by statistical binning process with the derivative spectra generated from the average of a large number of Brownian trajectories, are compared and yield a very good agreement. It is suggested that this method can be implemented to calculate absorption spectra from molecular dynamics (MD) simulation data. One of its advantages is that due to its reduction of computational effort, the parametrization process will be quicker. Secondly, the transition matrix defined in this manner, may indicate separable potential changes during the motion of the molecule and may have advantages when working with reducible set of coordinates. Thirdly, one can calculate the ESR spectra from a single MD trajectory directly without extending it artificially in the time axis. However, for short MD trajectories, the required statistical information can not be obtained depending on the timescale of transitions. Therefore, some statistical improvement will be needed in order to reach a better convergence.
\\

Keywords: Electron Spin Resonance, Markov State Model,  MD simulations

\end{abstract}

    \maketitle
    \def\s{\rule{0in}{0.28in}}
    \setlength{\LTcapwidth}{\columnwidth}

\section{Introduction}
Spin labeling theory \cite{Spinlabelingbook,SpinlabelingSc,Spinlabeling} has many applications in understanding the dynamics of complex molecules in a liquid environment. The line shapes from the continous wave electron spin resonance (cw-ESR) spectroscopy due to rotational diffusive motion in liquid environment had already been investigated by stochastic Kubo-Anderson approach \cite{Anderson,Kubo54,Kubo69}, approximation of relaxation times\cite{Redfield,Abragam}, and spherical Stochastic Liouville Equation (SLE) \cite{SchneiderFreed} formalism.
 
The work by Robinson and co-workers \cite{Robinson92} changed the perspective on looking to the problem by taking the rotational molecular trajectory as given. This has opened a new era, when the community started to infer possible interactions in the liquid environment that are effective on the cw-ESR spectrum by using molecular dynamics (MD) simulations following Steinhoff and Hubbell \cite{SteinhoffHubbell,Steinhoffetal2000}. This framework is based on calculating cw-ESR spectra of a spin label on a rotationally diffusing molecule from the motion of the three Euler angles $(\phi,\theta,\psi)$ that represent the orientational dynamics, and can be modeled with isotropic and anistropic rotational diffusion processes. 

 On the other hand, the diffusion of three Euler angles may depend on the internal dynamics and structural properties of the molecule and the spin label \cite{Liang,cytochrome,Fourthdim}. Thus, ESR spectroscopy is very helpful method in understanding the physics of such complex structures \cite {Cekan, Bennati,Maragakis,Columbus}. With the inclusion of the internal dynamics, this problem becomes more complex so that it requires additional set of variables. As a result modeling spin dynamics with sampling of different types of potentials \cite{Stoica,Structurebased,ConformSlow2007} has become necessary. In the end, one can combine internal dynamics obtained from MD data and the global diffusion of the molecule to account for the entire dynamics of a spin-labeled molecule \cite{DeSensi}.

In the recent years, Markov State Models (MSM) of conformational dynamics have been suggested, in order to interpret the slowly varying potential changes applied on the molecule or the spin label itself due to the effects of internal degrees of freedom \cite{SezerMarkov,Sezerbook}. This framework has been established on determining transition probabilities from relaxation timescales of each dynamical mode. MSM technique shows a promising scheme in understanding the features of intra- and inter-molecular phenomena \cite{Schwantes2014,McGibbon} in non-equilibrium dynamics.

Another difficulty for the determination of the effects resulting from spatial dynamics of a molecule from a MD trajectory, is that MD trajectories are often too short and would require to complete the rest of trajectory artificially, e.g. adding the paths together back and forth. To overcome this issue, Oganesyan suggested a novel technique to get an overall scheme for simulation from a truncated trajectory \cite{Oganesyan2007, Oganesyan2009, Oganesyan2011}. Another approach suggests calculating ESR spectra by using spherical SLE formalism and taking rotational diffusion parameters from the MD simulations \cite{Budil2006} which has basic similarities with the framework that we use in this study.

In this paper, we take the example of the cw-ESR spectrum of a spin-1/2 electron coupled to a magnetic field and spin-1 nucleus, e.g. nitroxide spin label, freely diffusing in a liquid, which is a problem that has been discussed many times. The s-state Kubo-Anderson process, with Markovian jumps maintains a solution that takes relatively less computational effort than working on the continuous coordinate system. We show that a Brownian trajectory can be mapped to a MSM by using a statistically binning process of exhibited jumps from one microstate to another. Therefore, a comparison between, on the one hand, calculated derivative spectra from MSM model with transition matrix obtained from a single Brownian trajectory by statistical binning process and, on the other hand the derivative spectra generated from the average of a large number of Brownian trajectories has been made and shows a very good agreement. In addition to that, extension to other coordinates, i.e., other rotation angles and hidden dynamics is noted. It is believed that this framework would fit into calculating  the derivative spectra MD simulation trajectories. This methodology can be also successful for separable and reducible potentials. It also allows a faster parametrization process with less computational time by reducing the set of variables. 

This article is organized as follows: The Kubo-Anderson process with s-state Markov model, generalization to s-state Kubo-Anderson process with discrete rotational diffusion, and continuous isotropic diffusion processes are overviewed in Sections 2, 3, and 4 respectively. In Section 5, a reduction scheme from continuous coordinates to discrete coordinates by using statistical binning process is introduced, and put to a test for the nitroxide example. In Section 6, we note an extension scheme to other coordinates and also exhibit the relevant comparisons. The application on short length trajectories is shown in Section 7. Finally, we discuss modeling different diffusion processes in Markov dynamics. In addition to that, we have included calculation tools for trajectory, evolution of average magnetization in trajectory method, and computational details in Appendices A,B, and C.

\section{MARKOV STATE MODEL FOR DIFFUSION}
For a Markov process, the master equation for microstate probability vector $\bra{p(t)}$ is given as

\begin{equation}
\label{eq1}
	\frac{d\bra{p(t)}}{dt}=\bra{p(t)}\bold{K},
\end{equation}
\\where $\bold{K}$ is the transition rate matrix per unit time whose elements $\bold{K}_{ij}$ represents the transition rate from microstate $\bra{i}$ to $\bra{j}$, so that it is basically the diffusion operator. The time-dependent average magnetization is a function of these microstates, and it can be written as $M_{+}(t)=\left\langle M_+(t)  |  p(t)\right\rangle$, or equivalently $M_{+}(t)=\left\langle p(t) M_+(t)  |  1\right\rangle$. Thus, the time evolution of probability-weighted magnetization vector is given as a Kubo-Anderson process\cite{Anderson,Kubo54,Kubo69},

\begin{equation}
\label{eq2}
	\frac{d\bra{p(t) M_{+}(t)}}{dt}=\bra{p(t) M_+(t)}(-i\bold{\Omega}+\bold{K}),
\end{equation}
\\where $\bold{\Omega}$ is a diagonal matrix with the eigenfrequencies of the Hamiltonian. Here,  
$\bold{\Omega}$ is independent of time if the microstates are stationary. The decay rate of the average magnetization is given by the transition rate matrix. The cw-ESR signal is obtained from the Fourier transform of the time-dependent average magnetization. Taking the Fourier transform of both sides and multiplying by ket $\ket{1}_s=[1,1,1,...,1]^T$, which is a vector with length s (number of states), and applying the initial condition $M_+(0)=0$ for all microstates yields to the equation;

\begin{equation}
	{I(\omega)}=\bra{v_{eq.}}(i\bold{\Omega}-\bold{K}+i\omega \bold{I}_s)^{-1}\ket{1}_s.
\end{equation}
where $\bra{v_{eq.}}$ represents equilibrium probability vector and $\bold{I}_s$ is $s\times s$ identity matrix. This is the Stochastic Liouville Equation (SLE) formalism for ESR absorption lines in which the Liouvillian operator is taken as the eigenfrequency matrix $\bold{\Omega}$. The normalization factors are ignored for practicality since we normalize the spectrum by the maximum intensity value. The spin-spin relaxation rate can be included as an operator in the parenthesis. Taking the derivative with respect to frequency $\omega$ will give the derivative absorption spectra,
\begin{equation}
	\frac{\partial I(\omega)}{\partial \omega}=-i\bra{v_{eq.}}(i\bold{\Omega}-\bold{K}+(i\omega+\gamma_e T_2^L))\bold{I}_s)^{-2}\ket{1}_s,
\end{equation}
where ${T_2^L}=1/\gamma_e T_2$, with $T_2$ as spin-spin relaxation time.
Note that we should divide each term in the paranthesis by gyromagnetic ratio $\gamma_e$ in order to see the spectrum in Gauss (G) units. This formalism can be extended to s-site jump model, i.e., the s-state Kubo-Anderson process \cite{Blume} and Eq.(3) will be the solution for absorption spectra. 

\section{ISOTROPIC ROTATIONAL DIFFUSION IN DISCRETE FORM}
MSM model can be used for describing the rotational diffusion process for molecules by using a discretized form of diffusion equation \cite{Robinson92,Lenk,GordonMess}.
The Hamiltonian that we consider here is the case where we have stationary eigenkets, i.e., we neglect the $I_{+}$, $I_{-}$ terms,

\begin{equation}
	\frac{H(t)}{\gamma_{e}}=\omega_0 g_{zz}^{lab}(t)S_z+S_{z}A_{zz}^{lab}(t)I_z.
\end{equation}
in which we define $\omega_0=B_0/g_e$. The spin label is assumed to be rigidly fixed to a macromolecule that is freely diffusing in a solution. The components $g_{zz}^{lab}$ and $A_{zz}^{lab}$ are found by a transformation using Euler angles ($\phi,\theta,\psi$) on the diaganol matrix consisting of their  xx, yy, and zz components in the molecular frame. For the cases in which their plane components x and y are equal to each other, i.e., $A_{xx}=A_{yy}$ and $g_{xx}=g_{yy}$, the only relevant diffusion coordinate is the angle $\theta$. Their transformation is given as
\begin{eqnarray}
	g_{zz}^{lab}(t)=cos^{2}(\theta(t))g_{zz}^{mol}+sin^{2}(\theta(t))g_{xx}^{mol},\nonumber\\
         A_{zz}^{lab}(t)=cos^{2}(\theta(t))A_{zz}^{mol}+sin^{2}(\theta(t))A_{xx}^{mol}.
\end{eqnarray}
Note that, we may subtract the rotation independent part,  i.e., $tr(\bold{g})/3$, in the first line of Eq.(6) which will result in a shift in the frequency axis as $\omega\rightarrow \omega-\omega_0$.  
In the MSM model of rotational diffusion, $\theta$ is discretized according to the number of states s in the form $\theta_k=(k-\frac{1}{2})\Delta \theta$, where k=1,2,...,s and $\Delta \theta=\pi/s$. Following the arrangement of microstates, the purpose is to find the transition rate matrix $\bold{K}$ for the discretized $\theta$ coordinate. The calculation of the transition rate matrix from the discretized form of the rotational diffusion equation by using finite difference method is given in detail in Ref.\cite{Robinson92}. Here, we apply the same procedure with reflective boundary conditions, and use transition rate matrix $\bold{K}$ to solve Eq. (3), which takes much less time than taking average over various diffusion trajectories. The magnetization is a summation over different orientations of the nuclear spin $m=-1,0,1$, i.e., $M_{+}(t)=\Sigma_{m}M_{+}^{m}(t)/3$, where the magnetization components have their distinct time evolution $M^m_{+}(t)=\left\langle M^m_+(t)  |  p(t)\right\rangle$, and therefore it is convenient to solve their contribution on the absorption spectra $I(\omega)$ separately and add them together. As said before, the matrix $\bold{\Omega}$ is the eigenfrequency matrix, whose components are in this case
\begin{equation}
	\frac{h\nu^{m}}{\gamma_{e}}=\omega_0 g_{zz}^{lab}+mA_{zz}^{lab}.
\end{equation}
Finally, for isotoropic rotational diffusion, we have $\bra{v_{eq.}}=\bra{sin(\theta)}$ in Eq.(3).

\section{ISOTROPIC ROTATIONAL DIFFUSION IN CONTINUOUS FORM} 
Another approach to this problem suggests generating Gaussian or uniform diffusion trajectories using the relevant diffusion constant and implying equilibrium conditions \cite{SteinhoffHubbell}. Accordingly, the rotation angle $\theta$ is continous, i.e., we can reach any angle between 0 and $\pi$, but the time-axis is discretized. 

The diffusion equation using the Euler angles is solved by the It\^{o} process \cite{Ito,Brillinger}. The equation of motion for angle $\theta$ is given as

\begin{equation}
	\Delta\theta_n=\sigma \Delta X_n + \frac{\sigma^2}{2tan\theta_n},
\end{equation}
where n is the time step, $\Delta X_n$ is the Brownian with mean zero, and $\sigma=\sqrt{2D\Delta t}$. The second term in Eq.(9) is due to the potential maintaining the path being in spherical coordinates. When this process is carried out by using a finite $\Delta t$, it makes strong jumps near the boundaries $\theta=0$ and $\theta=\pi$, which would cause a random noise. In order to get rid of this problem, it is convenient to take $\Delta t$ as small as possible, but this takes more computational time. Alternatively, this problem is solved by using quaternion based Monte-Carlo approach \cite{DeSensi, Sezerquater}.

The absorption spectrum is calculated through the Fourier transform of the average magnetization, i.e.,

\begin{equation}
I(\omega)=\int_{0}^{T}\Sigma_{m}\left\langle M^m_+(t)  |  p(t)\right\rangle e^{-i\omega t} e^{-t/T_2}dt,
\end{equation}
where again the normalization factor for intensity is ignored. The contribution of each trajectory to the magnetization is done by introducing their initial configuration from the equilibrium distribution, i.e., $p_{eq.}(\theta)=sin(\theta)$.

\begin{figure}[H]
\centering

\includegraphics{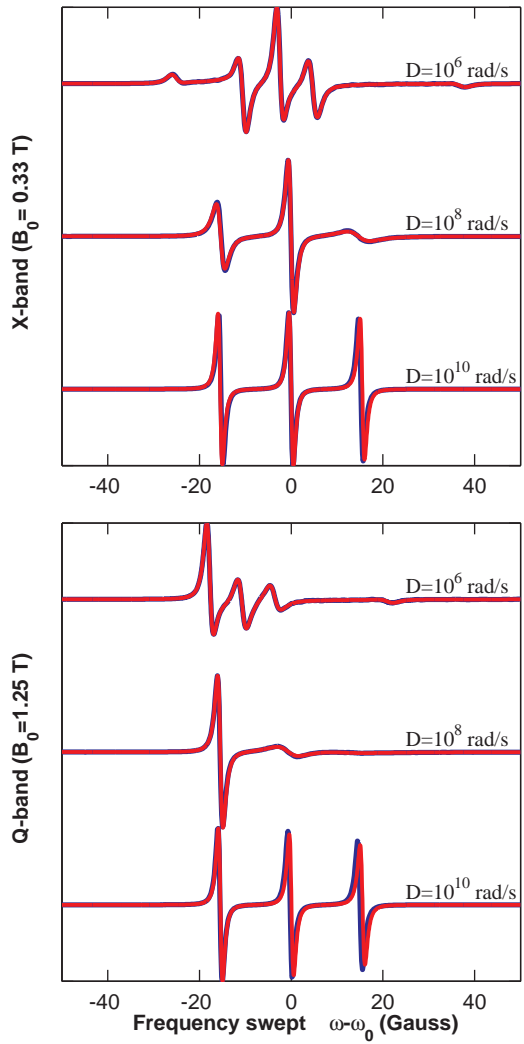}

\caption{Derivative spectra generated from 100,000 Brownian trajectories (blue line) with time step $\Delta t$=0.005 ns, $\Delta t$=0.025 ns, $\Delta t$=0.5 ns until magnetization significantly decays to zero, respectively $t=600 ns, t=1 \mu s, t=5 \mu s$ (and for the ones with $t<2$ $\mu$ s zero-padded to 2 $\mu$ s) and calculated from MSM model (red line) with s=12, 18, 36 states from bottom to top are compared. The two results are indistinguishable on the scale of this figure. Lorentzian broadening with $T_2^L$=0.8 G is used. The magnetic tensor parameters are given in Eqs. (10,11).}
\end{figure}

The comparison of the derivative spectra calculated from Brownian trajectories and MSM model is given in Fig. 1, to point out once more the equivalence of two methods. The magnetic tensor parameters are given as

\begin{equation}
(g_{xx},g_{yy},g_{zz})^{mol}=(2.00210,   2,00210,   2.00775),
\end{equation}
\begin{equation}
(A_{xx},A_{yy},A_{zz})^{mol}=(6.62,   6.62,   33.09)\ \ G .
\end{equation}

For the discrete jump model, the convergence is reached with just s=36,18,12 states for $D=10^6,10^8,10^{10}$ rad/s respectively. Thus, a lot of computational effort has been eliminated and minimized.

\section{REDUCTION FROM CONTINOUS SPACE TO DISCRETE COORDINATES} 
Reduction to finite number of microstates from the continous model is done first by creating the discretized angle axis with the angle set $\theta=\theta_1,\theta_2,...,\theta_s$, and then using binning procedure for the angle such that if angle $\theta(t)$ is between $\theta_k-\Delta \theta/2$  and $\theta_k+\Delta \theta/2$, it is set as $\theta(t)=\theta_k$ (Fig. 2). Accordingly, all the angles can be defined in such manner, and then we should be able to determine the transition probabilities between microstates.

\begin{figure}[h!]
\centering
\includegraphics{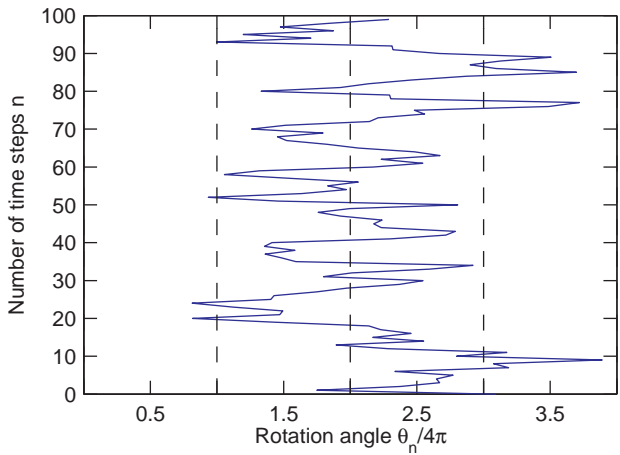}
\caption{Simple reduction scheme from continous to discrete coordinates for a Brownian trajectory for the s=4 MSM model. If a point in the trajectory falls into one of these gridlines, it takes the value of that bin. Calculation of the transition matrix is done using the discretized coordinates determined by this scheme.}
\end{figure}

The solution to Eq. (1) is simply:

\begin{equation}
	\bra{p(t)}=\bra{p(0)}\bold{U}(t)
\end{equation}
or
\begin{equation}
	\bra{p(t+\Delta t)}=\bra{p(t)}\bold{U}(\Delta t),
\end{equation}
where $\bold{U}(t)=e^{\bold{K}t}$ is the propagator for the microstate probability vector, or simply the transition probability matrix. In order to create the transition matrix from a Brownian diffusion trajectory, we should first be able to calculate the propagator matrix, take its matrix logarithm, divide by $\Delta t$. Our interest is in Eq. (14) and since we allow at most a single jump within a time step $\Delta t$ and only one component of $\bra{p(t)}$ is equal to 1 in a random trajectory, we may define $\bra{p(t)}=\bra{i}$ and $\bra{p(t+\Delta t)}=\bra{j}$, where $\bra{i}$  and $\bra{j}$ are chosen among the orthonormal eigenbasis of microstates. Multiplying both sides by ket $\ket{j}$ gives the matrix elements of $\bold{U}(\Delta t)$,

\begin{equation}
	1=\bra{i}\bold{U}(\Delta t)\ket{j}.
\end{equation}

Single jump at a time means that there is a contribution to a single component (i,j) of the matrix $\bold{U(t)}$ at each time step. Therefore, we should cover a long trajectory, and then take an average. The total probability of transitions from a state should be conserved, hence we should normalize the rows of the transition probability matrix $\bold{U(t)}$.

The timescale of events are governed by the rotational correlation time $\tau_c=1/(6D)$. The resolution of the transition probability matrix will be determined by the number of time steps we take into account, and typically a minimum of 10,000 time steps with a time step around $\Delta t \sim \tau_c/10$ is needed in order to have a reliable transition matrix. In this study, up to 40,000 time steps ($\Delta t$ values are given in the figures) from a single diffusion trajectory are taken into account, which is especially needed when approaching to the rigid limit, i.e., for $D=10^6$ rad/s in Fig. 3.

Finally, the equilibrium probability density vector can be determined from both transition matrix or occupancy rates. In this study we use the latter. Now we have all entries for Eq.(3) and Eq.(4) and can solve for the absorption spectrum. Alternatively, one can always calculate the average magnetization in time domain with the given ingredients. The comparison between the derivative spectra  generated from 100,000 Brownian trajectories and from MSM model with transition matrix obtained from a single Brownian trajectory with 40,000 time steps is given in Fig. 3. 

This approach can easily be applied to the calculation of absorption spectra from a single MD trajectory, and resolution of time steps can be taken same as the relevant MD simulation. In the next section, the extension to another rotation angle will be demonstrated before finally discussing the use for short MD trajectories and the possibilities of having a continuous spatial diffusion entangled to a m-state Markov model.

\begin{figure}
\centering
\includegraphics{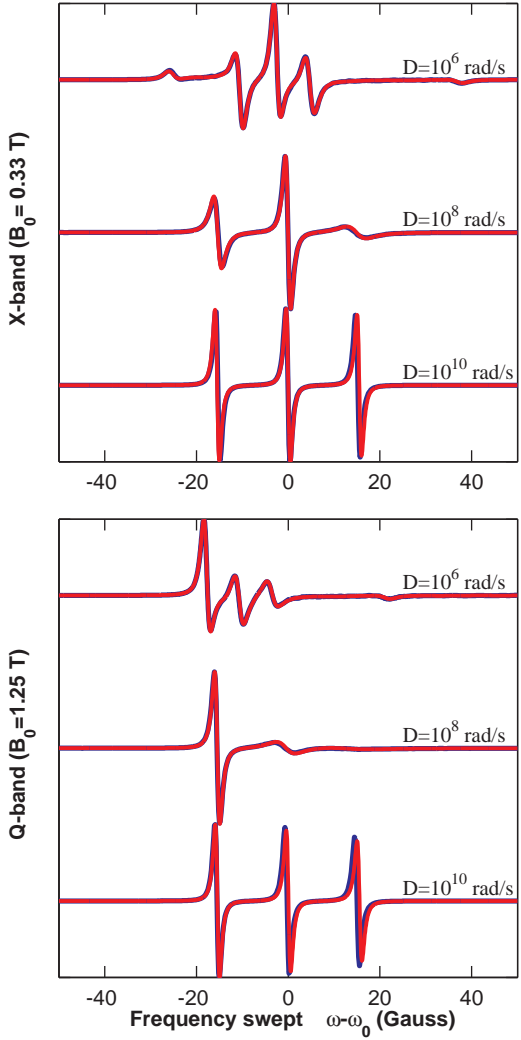}

\caption{Derivative spectra generated from 100,000 Brownian trajectories (blue line) same as in Fig. (1) and calculated from MSM model (red line) with transition matrix obtained from a single Brownian trajectory with 40,000 time steps having s=12, 18, 36 states from bottom to top are compared. The two results are indistinguishable on the scale of this figure. Lorentzian broadening with $T_2^L$=0.8 G is used. The magnetic tensor parameters are given in Eqs. (10,11).}
\end{figure}

\section{EXTENSION TO OTHER COORDINATES}
The diffusion operator that we want to consider is in the form 

\begin{equation}
	\Gamma(\theta,\phi)=D\frac{1}{sin\theta(t)}\frac{\partial}{\partial \theta}sin\theta(t)\frac{\partial}{\partial \theta} + \frac{D}{sin^2\theta(t)}\frac{\partial^2}{\partial\phi^2}.
\end{equation}

Let us define the transition rates for diffusion operator for angle $\theta$ with $s_1 \times s_1$ matrix $\bold{K}_{\theta}$ and for the operator $D\frac{\partial^2}{\partial\phi^2}$ with $s_2 \times s_2$  as $\bold{K}_{\phi}$ matrix, where $s_1$,$s_2$ are the numbers of eigenstates for angles $\theta$ and $\phi$ respectively. Hence, the transition rate  operator in Eq.(3) will be a $s_1 s_2\times s_1 s_2$ matrix

\begin{equation}
	\bold{K}_{\theta\phi}=\bold{K}_{\theta}\otimes \bold{I}_{s_2}+\bold{M}\otimes\bold{K}_{\phi}
\end{equation}
which is the discrete version of $\Gamma(\theta,\phi)$. $\bold{M}$ is a $s_1 \times s_1$ diagonal matrix with elements $1/sin^2(\theta_k)$ remembering $k=1,2,...,s_1$, and ${\bf I}_{s_2}$ is $s_2\times s_2$ identity matrix. The transition matrix in this case is expanded in $\ket{\theta, \phi}=\ket{\theta}_{s_1}\ket{\phi}_{s_2}$ space,
\begin{equation}
	\ket{\theta, \phi,m}=\ket{\theta}_{s_1}\ket{\phi}_{s_2}\ket{m}_3,
\end{equation}
where $\ket{\theta},\ket{ \phi}$ are respectively $s_1, s_2$ dimensional vectors, and $\ket{m}=[1,1,1]^T$. At equilibrium, the state vector becomes $\ket{sin\theta}\ket{1}\ket{1}$. Thus, it is the general formalism for an electron spin that is coupled to a magnetic field and nuclear spin, on a rotating frame with Euler angles ($\theta$,$\phi$). In discretized coordinates, these angles can be defined with two integers $(k_1,k_2)$ corresponding to the angle values $\theta_{k_1}=({k_1}-\frac{1}{2})\Delta \theta$ and $\phi_{k_2}=({k_2}-\frac{1}{2})\Delta \phi$ where $\Delta \theta=\pi/s_1$, $\Delta \phi=2\pi/s_2$ by following the reduction scheme in Fig. 2. For simplicity, we had eliminated the ket $\ket{m}$, in Section 3, by calculating a simple summation over m states.
The equation for absorption spectra in Eq.(3) becomes
\begin{equation}
	{I(\omega)}=\bra{v_{eq.}}(i \bold {\omega}+i {\bold{\cal L}}-\bold{K_{\theta\phi m}})^{-1}\ket{1}_{s_1\times s_2 \times 3},
\end{equation}
which will give the derivative absorption spectra as:
\begin{equation}
	\frac{\partial I(\omega)}{\partial \omega}=-i\bra{v_{eq.}}(i \bold {\omega}+i {\bold{\cal L}}-\bold{K_{\theta\phi m}})^{-2}\ket{1}_{s_1\times s_2 \times 3},
\end{equation}
where ${\bold{\cal L}}$ is the Liouvillian operator extended to $\theta$ and $\phi$ basis and $\bold{K_{\theta\phi m}}$ is the transition matrix extended to $\ket{m}$ basis, hence $\bold{K_{\theta\phi m}}=\bold{K_{\theta\phi}}\otimes \bold{I}_3$. The spin-spin relaxation rate which will result a Lorentzian broadening of $\gamma_e T_2^L$ in the spectrum, can be included in same manner as shown in Eq.(4). Note that we should divide each term in the paranthesis by gyromagnetic ratio $\gamma_e$ in order to see the spectrum in Gauss (G) units.

Including $I_{+}$, $I_{-}$ terms in Hamiltonian, we will have ${\cal H}=H(t)/\gamma_e$:
\begin{equation}
 {\cal H}=\omega_0g_{zz}^{lab} S_z  + S_z ( A_{zx}^{lab} I_x+ A_{zy}^{lab} I_y + A_{zz}^{lab} I_z)
\end{equation}
which can be represented in matrix form as:
\begin{equation}
{\cal H}=\frac{1}{2}
 \begin{pmatrix}
\begin{array}{c c c | c c c}
   & & & & &  \\
& {\cal H}^{\uparrow} & & & &  \\
& & & & &  \\
  \hline
   & & & & &  \\
& & & & {\cal H}^{\downarrow} &  \\
& & & & &  \\
   \end{array}
    \end{pmatrix}
\end{equation}
 where ${\cal H}^{\uparrow}=-{\cal H}^{\downarrow}$, and given as:
\begin{equation}
 {\cal H}^{\uparrow}=
 \begin{pmatrix}
 \omega_0 g_{zz}^{lab} + A_{zz}^{lab}  &  \frac{1}{\sqrt2} (A_{zx}^{lab}-iA_{zy}^{lab}) &  0 \\
 \frac{1}{\sqrt2} (A_{zx}^{lab}+iA_{zy}^{lab}) & \omega_0 g_{zz}^{lab}   &  \frac{1}{\sqrt2} (A_{zx}^{lab}-iA_{zy}^{lab}) \\
  0  &  \frac{1}{\sqrt2} (A_{zx}^{lab}+iA_{zy}^{lab}) &    \omega_0 g_{zz}^{lab} - A_{zz}^{lab}\\
 \end{pmatrix}.
 \end{equation}
 
 Liouvillian operator in this case is expanded in $\ket{\theta}_{s_1}\ket{\phi}_{s_2}\ket{m}_{s_3}$ basis such as for example when $s_1=3$,
 \begin{equation}
{\cal L}(\left \{ \theta \right \},\left \{ \phi \right \})=
 \begin{pmatrix}
\begin{array}{c c c | c c c | c c c}
   & & & & &  \\
& {\cal L} (\theta_1,\left \{ \phi \right \})& & & &  \\
& & & & &  \\
  \hline
   & & & & &  \\
& & & & {\cal L} (\theta_2,\left \{ \phi \right \}) &  \\
& & & & &  \\
  \hline
   & & & & &  \\
& & & & & & & {\cal L} (\theta_3,\left \{ \phi \right \}) &  \\
& & & & &  \\
   \end{array}
    \end{pmatrix}
 \end{equation}
 and  ${\cal L} (\theta_k,\left \{ \phi \right \})$ is extended in itself with the same hierarchy depending on the number of $\phi$ states, $s_2$.
We have concluded that the Hamiltonian in Eq. (21) is our relevant Liouvillian operator, i.e., ${\cal L}(\left \{ \theta \right \},\left \{ \phi \right \})={\cal H}^{\uparrow}(\left \{ \theta \right \},\left \{ \phi \right \})$ using the Hermiticity of both the Hamiltonian and the density matrix \cite{Landau} and after tracing with $S_+$ operator. The elements of ${\bf A}^{lab}$ and ${\bf g}^{lab}$ matrices are obtained as a function of three Euler angles $(\phi,\theta,\psi)$, defined with a transformation:

\begin{equation}
	{\bf A}^{lab}=R(\phi,\theta,\psi){\bf A}^{mol}R^T(\phi,\theta,\psi)
\end{equation}
where $R(\phi,\theta,\psi)$, is the rotation matrix from molecular frame to lab frame.

For creating the transition matrix $\bold{K}_{\theta\phi}$, the discrete form of the diffusion operator $\Gamma(\theta,\phi)$ in Eq.(15) which is defined in Eq.(16) may not represent successfully the spherical rotation, and instead of that we will use the methodology introduced in Section 5 extended to $\ket{\theta}_{s_1}\ket{\phi}_{s_2}$ space. Therefore we will first create a trajectory long enough which is generated by quaternion based Monte-Carlo algorithm, and then take the statistics from that trajectory by expressing the instantaneous state with $(k_1,k_2)$ using the information of both angles $\theta$ and $\phi$ at that time. Here, $(k_1,k_2)$ state in $\ket{\theta}_{s_1}\ket{\phi}_{s_2}$ eigenbasis can be expressed by the eigenket $\ket{\theta_{k_1}}_{s_1}\ket{\phi_{k_2}}_{s_2}$ whose all components are zero except the one at $((k_1-1)s_2+k_2)^{th}$ row which is equal to 1. 

The definition of Euler angles in terms of quaternions and their time evolution is explained in detail in Ref.\cite{DeSensi} and Ref.\cite{Sezerquater}, and also included in Appendix A. For the trajectory method, we need to keep track of evolution of density matrix and calculate average magnetization as a function of time from various trajectories, and then take the Fourier transform to obtain spectrum as shown in Eq. (9). The evolution of transverse magnetization in trajectory method is explained in Appendix B. The comparison between two methods for isotropic rotation with angles $(\theta,\phi)$,  using magnetic tensor parameters in Eqs. (10,11) is shown in Fig. (4) and yield a very good agreement.

\begin{figure}[h!]
\centering
\includegraphics{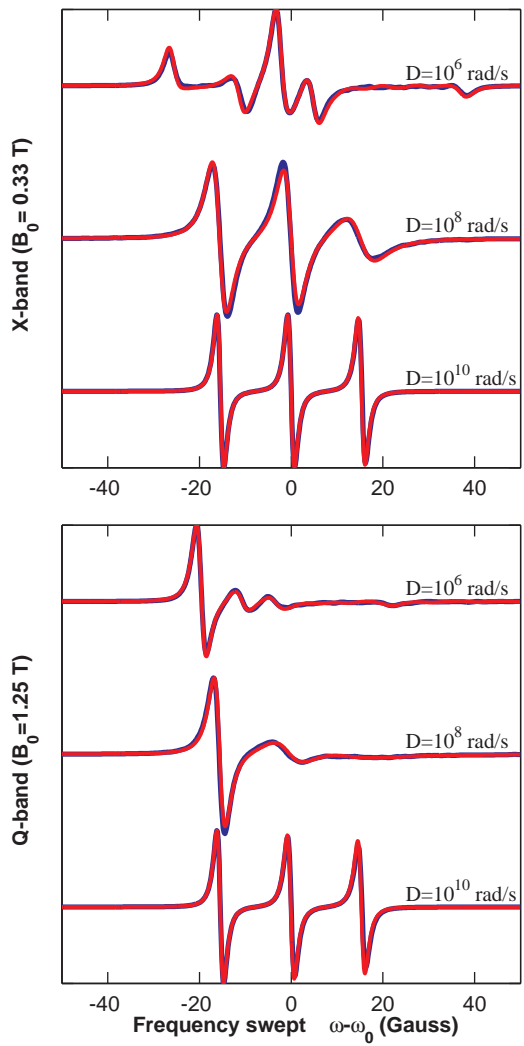}
\caption{Derivative spectra generated from 20,000 Brownian trajectories (blue line) with time step $\Delta t$=0.010 ns, $\Delta t$=0.2 ns, $\Delta t$=0.5 ns until magnetization significantly decays to zero, respectively $t=400 ns, t=700 ns, t=2.5 \mu s$ (and for the ones with $t<2$ $\mu$s zero-padded to 2 $\mu$s) and calculated from MSM model (red line) of single Brownian trajectory with 40,000 timesteps having $\Delta t$=0.005 ns,  $\Delta t$=0.2 ns, $\Delta t$=25 ns with $(s_1,s_2)$=(12,5) (18,5) (21,5) states from bottom to top are compared. The two results are indistinguishable on the scale of this figure. Lorentzian broadening with $T_2^L$=1.25 G is used. The magnetic tensor parameters are given in Eqs. (10,11).}

\end{figure}
Using the procedure applied in Eqs. (18-23) and explained above, it is straightforward to extend angle space into $\ket{\theta, \phi, \psi}=\ket{\theta}_{s_1}\ket{\phi}_{s_2}\ket{\psi}_{s_3}$ space. The comparison between two methods for isotropic rotation with angles $(\theta,\phi,\psi)$,  using magnetic tensor parameters $(g_{xx},g_{yy},g_{zz})^{mol}=(2.0082, 2.0060, 2.0023)$ and $(A_{xx},A_{yy},A_{zz})^{mol}=(7.0, 6.0, 36.0) G$ is shown in Fig. (5) and resulted a very good agreement. We have also made the comparison for fully anisotropic diffusion of cases $D_z > D_y > D_x$ and $D_y > D_x > D_z$ in Fig. (6), where $D_x, D_y, D_z$ are respectively diffusion constants for rotation around x, y, z axes. The results seem to deliver a good result but need improvement whether by increasing the number of events in single Brownian trajectory or increasing the number of states and/or developing the binning method. It is also observed that convergence to the spectra obtained from 25,000 Brownian trajectories is mostly and highly dependent on the accuracy of motional statistics along $\theta$ axis.

\begin{figure}[h!]
\centering
\includegraphics{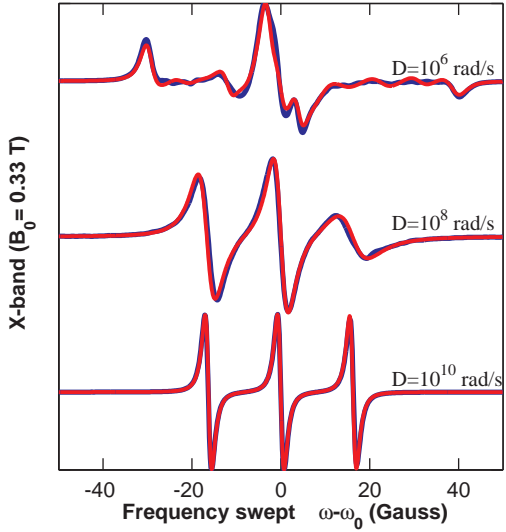}

\caption{Derivative spectra generated from 20,000 Brownian trajectories (blue line) with time step $\Delta t$=0.010 ns, $\Delta t$=0.2 ns, $\Delta t$=0.5 ns until magnetization significantly decays to zero, respectively $t=400 ns, t=700 ns, t=2.5 \mu s$(and for the ones with $t<2$ $\mu$s zero-padded to 2 $\mu$s) and calculated from MSM model (red line) of single Brownian trajectory with 40,000 time steps having $\Delta t$=0.005 ns,  $\Delta t$=0.2 ns, $\Delta t$=25 ns with $(s_1,s_2,s_3)$=(12,3,2) (18,3,2) (12,3,5) states from bottom to top are compared. The two results are indistinguishable on the scale of this figure. Lorentzian broadening with $T_2^L$=1.25 G is used. The magnetic tensor parameters are given as $(g_{xx},g_{yy},g_{zz})^{mol}=(2.0082, 2.0060, 2.0023)$ and $(A_{xx},A_{yy},A_{zz})^{mol}=(7.0, 6.0, 36.0) G$.}

\end{figure}

\begin{figure}[h!]
\centering
\includegraphics{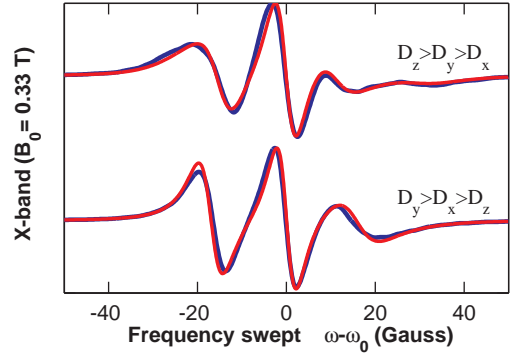}

\caption{Derivative spectra generated from 25,000 Brownian trajectories (blue line) for fully anisotropic rotational motion with time step $\Delta t$=0.2 ns until 200 ns for each trajectory (then, zero-padded to 2 $\mu$s),  and calculated from MSM model (red line) of single Brownian trajectory with 40,000 time steps having $\Delta t$=0.2 ns are compared. In the upper figure, we have $D_x=D/5, D_y=D/2, D_z=D$ using MSM with $(s_1,s_2,s_3)$=(30,3,2) while in the lower figure, we have $D_x=D/2, D_y=D, D_z=D/5$ using MSM with $(s_1,s_2,s_3)$=(36,5,2). For both cases, $D=10^8$ rad/s and Lorentzian broadening with $T_2^L$=1.8 G is used. The magnetic tensor parameters are same as in Fig. (5).}

\end{figure}

\section{APPLICATION TO SHORT BROWNIAN TRAJECTORIES}
In this section, our purpose is to discuss the applicability of the presented approach to short MD simulations. As a specific case, we will continue with isotropic rotational diffusion up to 100 ns. In that case, we would not expect to get the correct results for the rare event case, i.e., rigid limit, $D=10^6$ rad/s. Therefore, for such examples, it is better to estimate rotational correlation time, and perform simulations. On the other extreme, we expect and obtain a perfect convergence (data not shown) for fast motion limit, i.e., $D=10^{10}$ rad/s for all independent single trajectories. Our interest will be on the case of slow motional regime having $D=10^8$ rad/s. The procedure is the same as in Section 5 and 6, except that this time we have a Brownian trajectory of 100 ns. 

\begin{figure}[H]
\centering
\includegraphics{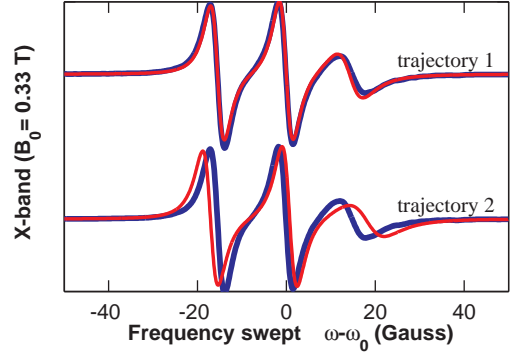}

\caption{Derivative spectra for $D=10^8$ rad/s calculated from MSM model (red line) of single Brownian trajectory until 100 ns having $\Delta t$=0.1 ns, with $(s_1,s_2)$=(18,5) states. Upper and lower figures are for different realizations of the single Brownian trajectory, blue lines are for comparison with the derivative spectra generated from 20,000 Brownian trajectories (blue lines) with time step $\Delta t$=0.2 ns until 700 ns as in Fig. (4). Lorentzian broadening with $T_2^L$=1.25 G is used. The magnetic tensor parameters are given in Eqs. (10,11).}
\end{figure}
For the diffusion process with $D=10^8$ rad/s the rotational correlational time is $\tau_c\approx1.667$ ns. Thus, the motion will still not reach some regions in phase space. Accordingly, when creating the transition probability matrix we may observe unvisited sections. It is important to eliminate rows with zero event, not only for computational ease but also not to get a singular matrix. As a result, for example, while starting with $s=18$ state model, we may end up with a $14\times14$ matrix instead of $18\times18$. This also shows that some statistical improvement is necessary when creating the gridlines. In our calculations, we have just followed the scheme that is creating gridlines equal in length. As seen in Fig. 7, the convergence of the derivative spectra calculated from MSM of single short Brownian trajectory to the one obtained from 20,000 trajectories may depend on the realization. In that case, we may need more information about the type of motion. This is shown in Fig. 8 with the same MSM having a transition matrix obtained from 5 independent trajectories, and it displays a better agreement. The convergence to the one obtained from 20,000 trajectories increases further with higher number of $\theta$ states.

\begin{figure}[H]
\centering

\includegraphics{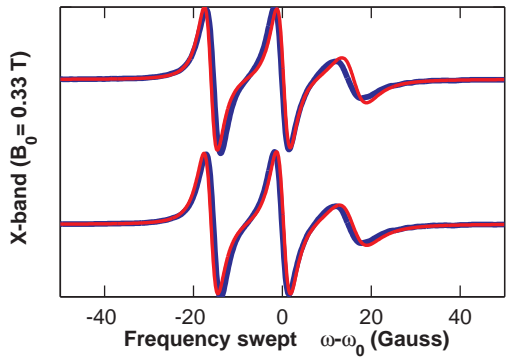}

\caption{Derivative spectra for $D=10^8$ rad/s calculated from MSM model (red line) of statistics obtained from 5 independent Brownian trajectories until 100 ns having $\Delta t$=0.1 ns. Upper figure is obtained from $(s_1,s_2)$=(18,5) state model and lower figure is obtained by increasing number of states, i.e., $(s_1,s_2)$=(32,5), and shows a better agreement, blue lines are for comparison with the derivative spectra generated from 20,000 Brownian trajectories (blue lines) with time step $\Delta t$=0.2 ns until 700 ns as in Fig. (4). Lorentzian broadening with $T_2^L$=1.25 G is used. The magnetic tensor parameters are given in Eqs. (10,11).}
\end{figure}

\section{DISCUSSION AND CONCLUSIONS}
As a further step, having a process such as spatial diffusion of molecule under a potential that is an element of m-state potential space in which the potential is selected by a Markov process, may be helpful in understanding the effect of conformational changes in the spin dynamics \cite{Sezerbook}. For example, a diffusive process that has a m-fold selective potential for the angle $\theta$ can be introduced, such as with m=A state for $\theta\in[0,\pi/2]$ and m=B state $\theta\in[\pi/2,\pi]$. This will help to get a quick feedback about the credibility of the model for the given MD simulation results by reducing the set of variables. If the simulations are done distinctly for two different types of potentials {\cite{Stoica,Structurebased}, then the overall transition rate matrix can be created by their extension to the potential subspace, i.e., taking their kronecker product with potential change rate matrix, and summing them up. Alternatively, if these two type of potentials are completely separable, then the two simulations can be defined in one transition probability matrix, by adding their records of jumps as if they are in the same trajectory, one being in the starting section, and the other in the second section, and their own lengths in time axis will be directly related to their equilibrium distribution and transition rates among these two types of potentials. On the other hand, for slowly varying potentials due to internal dynamics, this approach will be insufficient for not recognizing non-Markovian processes and therefore one has to model rotational motion of the molecule entangled with the effects of the conformational changes on the spin label itself, as defined in the upper statement and previous examples \cite{Sezerbook,Oganesyan2007}. 

In summary, we have simulated the cw-ESR spectrum of a spin-1/2 electron coupled to a magnetic field and spin-1 nucleus for X- and Q- bands by using both discrete isotropic rotational diffusion and continuous Brownian diffusion processes. In addition to that, calculated derivative spectra from the MSM model with transition matrix obtained from a single Brownian trajectory by statistical binning process and the spectra generated from the average of a large number of Brownian trajectories are compared and resulted in a very good agreement. It is suggested that this method can be implemented to calculate absorption spectra from MD simulation data. One of its advantages is that due to its reduction of computational effort, the parametrization process will be quicker. Secondly, the transition matrix defined in this manner, may indicate separable potential changes during the motion of the molecule. It is also possible to change the course of single trajectory by hand and observe its consequences. Thirdly, one can calculate the ESR spectra from a single MD trajectory directly without extending it artificially in the time axis. However, for short MD trajectories, the required statistical information can not be obtained depending on the timescale of transitions. Therefore, some statistical improvement will be needed in order to reach a better convergence. On the other hand, if there is a requirement of high precision in the extended coordinates, it will enlarge the transition matrix and therefore will take more computational time. Hence, reducibility of relevant coordinates is undoubtably necessary to maintain the efficiency of the present framework. It is hoped that, in the following studies, this framework will be helpful in extracting the statistics of motional information of a molecule or implementing additional motional information on the spin label under various kinds of potentials.

\section{ACKNOWLEDGEMENTS}
I would like to thank D. Sezer for bringing this problem to my attention and helpful discussions. I would also like to thank A. N. Berker for a critical reading.
 \appendix
 \section{Quaternion dynamics}
 For the generation of Brownian trajectories, we have used the conjecture in Ref. \cite{DeSensi} in which the detailed calculations can be found. Only difference is that here we use Gaussian random processes, instead of random uniform displacements. Euler angles $(\phi,\theta,\psi)$ defined in terms of quaternions as:
   \begin{equation}
{\bf q(t)}=
 \begin{pmatrix}
\begin{array}{c}
q_0(t)\\
q_1(t)\\
q_2(t)\\
q_3(t)
   \end{array}
    \end{pmatrix}
    = \begin{pmatrix}
\begin{array}{c}
cos\frac{\theta(t)}{2}cos\frac{(\phi(t)+\psi(t))}{2}\\
sin\frac{\theta(t)}{2}sin\frac{(\phi(t)-\psi(t))}{2}\\
-sin\frac{\theta(t)}{2}cos\frac{(\phi(t)-\psi(t))}{2}\\
-cos\frac{\theta(t)}{2}sin\frac{(\phi(t)+\psi(t))}{2}
   \end{array}
    \end{pmatrix}
 \end{equation}
 satisfying $q^2_0(t)+q^2_1(t)+q^2_2(t)+q^2_3(t)=1$, therefore it is still a function of 3 independent variables. Using the equation relating angular velocities around $x, y, z$ axes with the time derivative of Euler angles and converting it to quaternion formalism, one would get the equation of motion as:
   \begin{equation}
   {\bf q}(t+\Delta t)=e^{{\bf P}\Delta t/2}{\bf q}(t)
   \end{equation}
   where ${\bf P}$ is defined in the form of

  \begin{equation}
{\bf P}=
 \begin{pmatrix}
\begin{array}{c c c c}
0 & \omega_x & \omega_y & \omega_z \\ 
-\omega_x & 0 & \omega_z & -\omega_y \\ 
-\omega_y & -\omega_z & 0 & \omega_x \\ 
-\omega_z & \omega_y & -\omega_x & 0
   \end{array}
    \end{pmatrix}.
 \end{equation}

$\omega_x$, $\omega_y$, $\omega_z$ are respectively angular velocities around $x, y, z$ axes. As a realization of Brownian trajectory, we can evaluate  ${\bf P}\Delta t$ with 3 independent random Gaussian displacements. Thus, we introduce $\omega_i\Delta t=\sigma_i\Delta U_i$, where $\Delta U_i$ is a Brownian with mean zero and $\sigma_i=\sqrt{2D_i\Delta t}$, with $D_i$ as diffusion constant for rotation around $i^{th}$ axis. Simplifications for the time evolution of quaternions is shown in the related reference. 

Rotation given by three Euler angles defined with the rotation matrix $R(\phi,\theta,\psi)=R_z(\psi)R_x(\theta)R_z(\phi)$ which can be written in terms of quaternions as

\begin{widetext}
  \begin{equation}
R(t)=
 \begin{pmatrix}
\begin{array}{c c c}
q^2_0+q^2_1-q^2_2-q^2_3 & -2(q_0q_3-q_1q_2) & 2(q_0q_2+q_1q_3) \\ 
2(q_0q_3+q_1q_2) & q^2_0-q^2_1+q^2_2-q^2_3 & -2(q_0q_1-q_2q_3) \\
-2(q_0q_2-q_1q_3) & 2(q_0q_1+q_2q_3) & q^2_0-q^2_1-q^2_2+q^2_3
   \end{array}
    \end{pmatrix}.
 \end{equation}
 \end{widetext}

Accordingly, one can use directly the rotation matrix in this form to implement Eq. (24) and calculate the Hamiltonian. In our conjecture for calculating the transition matrix from discretized coordinates, we take the information by converting quaternions to three Euler angles at each time step throughout the trajectory. It is also possible to consider an application of binning process over quaternions which are taking values between -1 and 1 with the condition $q^2_0(t)+q^2_1(t)+q^2_2(t)+q^2_3(t)=1$, and therefore belonging to a process of 3 independent variables. In this study, we have  implemented binning process only on to angle values.

 \section{Propagation of magnetization}
The time dependent transverse magnetization observable is obtained as:
\begin{equation}
M_+=Tr( \rho S_+).
\end{equation}
As our Hamiltonian shown in Eq. (5) and Eqs. (20-22), we will have a $6\times6$ density matrix. $S_+$ operator in that basis will act on only $\rho_+$ section of density matrix which is shown in $\rho$ as,
   \begin{equation}
   \rho=
 \begin{pmatrix}
\begin{array}{c | c c}
 & &   \\  \hline
 \rho_+ & & 
   \end{array}
    \end{pmatrix}
 \end{equation}
 and hence $M_+=Tr(\rho_+)$. As a consequence we will need only the time evolution of $\rho_+$ to calculate time dependent magnetization. In short time dynamics, i.e., for small $\Delta t$ its time evolution can be calculated with \cite{Sezerquater}
 \begin{equation}
 \rho_+(t+\Delta t)=e^{i(\gamma_e{\cal H}^{\uparrow}){\Delta t}}\rho_+(t)e^{i(\gamma_e{\cal H}^{\uparrow}){\Delta t}}.
 \end{equation}
 Finally, initial condition of $\rho_+$ being a $3\times3$ identity matrix, i.e., $\rho_+(0)={\bf I}_3$ is implemented and propagation of that matrix is followed with its trace being recorded at each time step throughout the trajectory. Average magnetization that is needed to solve Eq. (9) is calculated over different realizations of Brownian diffusion with their appropriate $sin(\theta)$ weight. 
 
\section{Computational details}
The calculations are performed by MATLAB R2012b using a single core Intel\textsuperscript{\circledR} Xeon\textsuperscript{\circledR} Processor E5420 with base frequency at 2.50 Ghz.  
\subsection{Trajectory method}
Calculations are done with a simple process consisting of loop for time steps $n_{time}$, inside a loop for number of trajectories $N_{traj}$ in which the magnetization is determined as shown in the text and Appendix B. For the evaluation of exponentials in Eq. (A2) and (B3), we have used the simplifications suggested in the related references. To obtain the derivative spectrum, a Fast Fourier Transform is held using fft function in MATLAB. 

For the calculation of average magnetizations it is seen that there is a linear increase with number of steps as well as with number of trajectories. For ex., from average over 5 computations we have observed that for diffusion of $(\theta,\phi)$ with 5,000 time steps ($n_{time}$) and 10,000 Brownian trajectories ($N_{traj}$) average run time is $t_{av}=12257.6 s$ with a standard deviation $\sigma_t=85.48 s$, whereas with $n_{time}$=5,000 and $N_{traj}$=5,000, we had $t_{av}=6080.5 s$ and  $\sigma_t=28.51 s$, with $n_{time}$=5,000 and $N_{traj}$=1,000, we had $t_{av}=1223 s$ and  $\sigma_t=10.06 s$, and finally with $n_{time}$=1,000 and $N_{traj}$=1,000, we had $t_{av}=244.6 s$ and  $\sigma_t=1.07 s$. In addition to that, we have observed computation values for the case of diffusion of  $(\theta,\phi,\psi)$ being very similar values such as with $n_{time}$=1,000 and $N_{traj}$=1,000, $t_{av}=258.4 s$ and  $\sigma_t=1.97 s$. Accordingly, the computation values for each case obtained in this study can be determined from the relation  $t_{av}\propto N_{traj}n_{time}$, and using  

   \begin{gather*}
 (\theta) diffusion\nonumber\\
\Delta t=0.005 ns \longrightarrow 500 ns \longrightarrow 100,000 steps\nonumber\\
\Delta t=0.025 ns \longrightarrow 1 \mu s \longrightarrow 40,000 steps\\
\Delta t=0.5 ns \longrightarrow 5 \mu s \longrightarrow 10,000 steps
\end{gather*}

   \begin{gather*}
 (\theta, \phi) diffusion\nonumber\\
\Delta t=0.010 ns \longrightarrow 400 ns \longrightarrow 40,000 steps\nonumber\\
\Delta t=0.2 ns \longrightarrow 700 ns \longrightarrow 3,500 steps\\
\Delta t=0.5 ns \longrightarrow 2 \mu s \longrightarrow 4,000 steps
\end{gather*}

   \begin{gather*}
 (\theta, \phi, \psi) diffusion\nonumber\\
\Delta t=0.010 ns \longrightarrow 400 ns \longrightarrow 40,000 steps\nonumber\\
\Delta t=0.2 ns \longrightarrow 700 ns \longrightarrow 3,500 steps\\
\Delta t=0.5 ns \longrightarrow 2.5 \mu s \longrightarrow 5,000 steps.
\end{gather*}

It should be noted that the computation times for only $(\theta)$ diffusion are much lower than for the given examples, and it is completed within a few minutes or less while it takes less than 10 seconds for MSM method. Furthermore, convergence is reached with around 5,000-10,000 trajectories for $(\theta)$ diffusion and with more than 10,000 trajectories for two and three angles diffusions. Another point is that one can also reduce computation times by changing the total number of time steps (depending on additional broadenings) and time step value $\Delta t$, e.g., for fast motional limit, i.e. $D=10^{10}$ rad/s, one can get the same results(data not shown) with $\Delta t=50 ps$.

After calculation of average magnetization, Fast Fourier Transform is performed using fft function in MATLAB, which is completed within seconds for up to 10,000 time steps (including zero-padding), and depending on the size of time array the computation time may extend drastically. 
\subsection{MSM model of single Brownian trajectory}
The computation of derivative spectra from MSM of a single trajectory consists of $i)$ generation of single Brownian trajectory $ii)$ creation of transition matrix as explained in Section 5, $iii)$ calculation of energies for allowed states (depending on the number of states), $iv)$ calculation of derivative spectra using formula in Eq. (19) for 796 points between frequencies $\omega-\omega_0=[-50,50]$ G. Computation times for derivative spectra in the case of $(\theta,\phi)$ and $(\theta,\phi,\psi)$ diffusion which are exhibited in Fig. 4 and Fig. 5, are shown respectively in Table C.1 and Table C.2.
    \begin{table}[h!]
        \begin{tabular}{c c c | c c c | c c c}
            \hline
 & States (s1,s2) &  & & $t_{av}$ & &  & $\sigma_t$ &\\
  \hline
  & (12,5) &  & & 7.29 s& &  & 0.042 s&\\
 \hline
  & (18,5) &  & & 21.42 s& &  & 0.269 s&\\
 \hline
  & (21,5) &  & & 32.83 s& &  & 0.086 s&\\
 \hline
        \end{tabular}
\caption{ Average computation times $t_{av}$ and standard deviation $\sigma_t$ over 5 runs for calculation of spectra with $(\theta,\phi)$ diffusion. }
        \label{tab:1}
    \end{table}
    
        \begin{table}[h!]
        \begin{tabular}{c c c | c c c | c c c}
            \hline
 & States (s1,s2,s3) &  & & $t_{av}$ & &  & $\sigma_t$ &\\
  \hline
  & (12,3,2) &  & & 11.82 s& &  & 0.073 s&\\
 \hline
  & (18,3,2) &  & & 36.01 s& &  & 0.357 s&\\
 \hline
  & (12,3,5) &  & & 43.48 s& &  & 0.922 s&\\
 \hline
        \end{tabular}
\caption{Average computation times $t_{av}$ and their standard deviation $\sigma_t$ over 5 runs for calculation of spectra with $(\theta,\phi,\psi)$ diffusion.}
        \label{tab:2}
    \end{table}
For much larger matrices, this method becomes problematic, and decrease the computer efficiency drastically, with taking the logarithm and then the inverse of large matrices. In such cases, better optimizations and algorithms will be needed to improve this methodology. 

\bibliography{ESRfromMSMv2arx}

\begin{thebibliography}{10}
\expandafter\ifx\csname url\endcsname\relax
  \def\url#1{\texttt{#1}}\fi
\expandafter\ifx\csname urlprefix\endcsname\relax\def\urlprefix{URL }\fi
\expandafter\ifx\csname href\endcsname\relax
  \def\href#1#2{#2} \def\path#1{#1}\fi

\bibitem{Spinlabelingbook}
J.~L. Berliner, Spin Labeling: Theory and Applications, Academic Press, 1979.

\bibitem{SpinlabelingSc}
P.~P. Borbat, A.~J. Costa-Filho, K.~A. Earle, J.~K. Moscicki, J.~H. Freed,
  Electron spin resonance in studies of membranes and proteins, Science 291
  (2001) 266--269.
\newblock \href {http://dx.doi.org/10.1126/science.291.5502.266}
  {\path{doi:10.1126/science.291.5502.266}}.

\bibitem{Spinlabeling}
J.~Klare, H.-J. Steinhoff, Spin labeling {EPR}, Photosynth. Res. 102 (2009)
  377--390.
\newblock \href {http://dx.doi.org/10.1007/s11120-009-9490-7}
  {\path{doi:10.1007/s11120-009-9490-7}}.

\bibitem{Anderson}
P.~W. Anderson, A mathematical model for the narrowing of spectral lines by
  exchange or motion, J. Phys. Soc. Jpn. 9 (1954) 316--339.
\newblock \href {http://dx.doi.org/10.1143/JPSJ.9.316}
  {\path{doi:10.1143/JPSJ.9.316}}.

\bibitem{Kubo54}
R.~Kubo, Note on the stochastic theory of resonance absorption, J. Phys. Soc.
  Jpn. 9 (1954) 935--944.
\newblock \href {http://dx.doi.org/10.1143/JPSJ.9.935}
  {\path{doi:10.1143/JPSJ.9.935}}.

\bibitem{Kubo69}
R.~Kubo, A stochastic theory of line shape, Adv. Chem. Phys. 15 (1969)
  101--127.
\newblock \href {http://dx.doi.org/10.1002/9780470143605.ch6}
  {\path{doi:10.1002/9780470143605.ch6}}.

\bibitem{Redfield}
A.~Redfield, On the theory of relaxation processes, IBM J. Res. Dev. 1 (1957)
  19--31.
\newblock \href {http://dx.doi.org/10.1147/rd.11.0019}
  {\path{doi:10.1147/rd.11.0019}}.

\bibitem{Abragam}
A.~Abragam, The Principles of Nuclear Magnetism, Oxford University Press, 1961.

\bibitem{SchneiderFreed}
D.~J. Schneider, J.~H. Freed, Spin relaxation and motional dynamics, Adv. Chem.
  Phys. 73 (1989) 387--527.
\newblock \href {http://dx.doi.org/10.1002/9780470141229.ch10}
  {\path{doi:10.1002/9780470141229.ch10}}.

\bibitem{Robinson92}
L.~J.~S. B.~H.~Robinson, F.~P. Auteri, Direct simulation of continuous wave
  electron paramagnetic resonance spectra from brownian dynamics trajectories,
  J. Chem. Phys. 96 (1992) 2609.
\newblock \href {http://dx.doi.org/10.1063/1.462869}
  {\path{doi:10.1063/1.462869}}.

\bibitem{SteinhoffHubbell}
H.-J. Steinhoff, W.~L. Hubbell, Calculation of electron paramagnetic resonance
  spectra from brownian dynamics trajectories: application to nitroxide side
  chains in proteins, Biophys. J. 71 (1996) 2201"1¤712.
\newblock \href {http://dx.doi.org/10.1016/S0006-3495(96)79421-3}
  {\path{doi:10.1016/S0006-3495(96)79421-3}}.

\bibitem{Steinhoffetal2000}
H.-J. Steinhoff, M.~Muller, C.~Beier, M.~Pfeiffer, Molecular dynamics
  simulation and epr spectroscopy of nitroxide side chains in
  bacteriorhodopsin, J. Mol. Liq. 84 (2000) 17--27.
\newblock \href {http://dx.doi.org/10.1016/S0167-7322(99)00107-5}
  {\path{doi:10.1016/S0167-7322(99)00107-5}}.

\bibitem{Liang}
Z.~Liang, Y.~Lou, J.~H. Freed, L.~Columbus, W.~L. Hubbell, A multifrequency
  electron spin resonance study of {T4} lysozyme dynamics using the slowly
  relaxing local structure model, J. Phys. Chem. B 108 (2004) 17649"1¤7659.
\newblock \href {http://dx.doi.org/10.1021/jp0484837}
  {\path{doi:10.1021/jp0484837}}.

\bibitem{cytochrome}
K.~Murzyn, T.~Rog, W.~Blicharski, M.~Dutka, J.~Pyka, S.~Szytula, W.~Froncisz,
  Influence of the disulfide bond configuration on the dynamics of the spin
  label attached to cytochrome c, Proteins 62 (2006) 1088--1100.
\newblock \href {http://dx.doi.org/10.1002/prot.20838}
  {\path{doi:10.1002/prot.20838}}.

\bibitem{Fourthdim}
H.~S. Mchaourab, P.~R. Steed, K.~Kazmier, Toward the fourth dimension of
  membrane protein structure: Insight into dynamics from spin-labeling {EPR}
  spectroscopy, Structure 19 (2011) 1549--1561.
\newblock \href {http://dx.doi.org/10.1016/j.str.2011.10.009}
  {\path{doi:10.1016/j.str.2011.10.009}}.

\bibitem{Cekan}
P.~Cekan, S.~T. Sigurdsson, Identification of single-base mismatches in duplex
  {DNA} by {EPR} spectroscopy, J. Am. Chem. Soc. 131 (2009) 18054"1¤7056.
\newblock \href {http://dx.doi.org/10.1021/ja905623k}
  {\path{doi:10.1021/ja905623k}}.

\bibitem{Bennati}
M.~Bennati, T.~F. Prisner, New developments in high field electron paramagnetic
  resonance with applications in structural biology, Rep. Prog. Phys. 68 (2005)
  411.
\newblock \href {http://dx.doi.org/10.1088/0034-4885/68/2/R05}
  {\path{doi:10.1088/0034-4885/68/2/R05}}.

\bibitem{Maragakis}
P.~Maragakis, K.~Lindorff-Larsen, M.~P. Eastwood, R.~O. Dror, J.~L. Klepeis,
  I.~T. Arkin, M.~O. Jensen, H.~Xu, N.~Trbovic, R.~A. Friesner, A.~G. Palmer,
  D.~E. Shaw, Microsecond molecular dynamics simulation shows effect of slow
  loop dynamics on backbone amide order parameters of proteins, J. Phys. Chem.
  B 112 (2008) 6155"1¤758.
\newblock \href {http://dx.doi.org/10.1021/jp077018h}
  {\path{doi:10.1021/jp077018h}}.

\bibitem{Columbus}
L.~Columbus, T.~Kalai, J.~Jeko, K.~Hideg, W.~L. Hubbell, Molecular motion of
  spin labeled side chains in $\alpha$-helices: analysis by variation of side
  chain structure, Biochemistry 40 (2001) 3828"1¤746.
\newblock \href {http://dx.doi.org/10.1021/bi002645h}
  {\path{doi:10.1021/bi002645h}}.

\bibitem{Stoica}
I.~Stoica, Using molecular dynamics to simulate electronic spin resonance
  spectra of {T4} lysozyme, J. Phys. Chem. B 108 (2004) 1771"1¤782.
\newblock \href {http://dx.doi.org/10.1021/jp036121d}
  {\path{doi:10.1021/jp036121d}}.

\bibitem{Structurebased}
C.~Beier, H.~J. Steinhoff, A structure-based simulation approach for electron
  paramagnetic resonance spectra using molecular and stochastic dynamics
  simulations, Biophys. J. 91 (2006) 2647--2664.
\newblock \href {http://dx.doi.org/10.1529/biophysj.105.080051}
  {\path{doi:10.1529/biophysj.105.080051}}.

\bibitem{ConformSlow2007}
D.~Hamelberg, C.~A.~F. de~Oliveira, J.~A. McCammon, Sampling of slow diffusive
  conformational transitions with accelerated molecular dynamics, J. Chem.
  Phys. 127 (2007) 155102.
\newblock \href {http://dx.doi.org/10.1063/1.2789432}
  {\path{doi:10.1063/1.2789432}}.

\bibitem{DeSensi}
S.~C. DeSensi, D.~P. Rangel, A.~H. Beth, T.~P. Lybrand, E.~J. Hustedt,
  Simulation of nitroxide electron paramagnetic resonance spectra from brownian
  trajectories and molecular dynamics simulations, Biophys. J. 94 (2008) 3798
  -- 3809.
\newblock \href {http://dx.doi.org/10.1529/biophysj.107.125419}
  {\path{doi:10.1529/biophysj.107.125419}}.

\bibitem{SezerMarkov}
D.~Sezer, J.~H. Freed, B.~Roux, Using markov models to simulate electron spin
  resonance spectra from molecular dynamics trajectories, J. Phys. Chem. B 112
  (2008) 11014"1¤7027.
\newblock \href {http://dx.doi.org/10.1021/jp801608v}
  {\path{doi:10.1021/jp801608v}}.

\bibitem{Sezerbook}
D.~Sezer, B.~Roux, Markov state and diffusive stochastic models in electron
  spin resonance, in: G.~Bowman, V.~Pande, F.~Noe (Eds.), An Introduction to
  Markov State Models and Their Application to Long Timescale Molecular
  Simulations, Springer, 2014.

\bibitem{Schwantes2014}
C.~R. Schwantes, R.~T. McGibbon, V.~S. Pande, Perspective: Markov models for
  long-timescale biomolecular dynamics, J. Chem. Phys. 141 (2014) 090901.
\newblock \href {http://dx.doi.org/10.1063/1.4895044}
  {\path{doi:10.1063/1.4895044}}.

\bibitem{McGibbon}
R.~T. McGibbon, C.~R. Schwantes, V.~S. Pande, Statistical model selection for
  markov models of biomolecular dynamics, J. Phys. Chem. B 118 (2014)
  6475"1¤781.
\newblock \href {http://dx.doi.org/10.1021/jp411822r}
  {\path{doi:10.1021/jp411822r}}.

\bibitem{Oganesyan2007}
V.~S. Oganesyan, A novel approach to the simulation of nitroxide spin label epr
  spectra from a single truncated dynamical trajectory, J. Magn. Reson. 188
  (2007) 196--205.
\newblock \href {http://dx.doi.org/10.1016/j.jmr.2007.07.001}
  {\path{doi:10.1016/j.jmr.2007.07.001}}.

\bibitem{Oganesyan2009}
V.~S. Oganesyan, E.~Kuprusevicius, H.~Gopee, A.~N. Cammidge, M.~R. Wilson,
  Electron paramagnetic resonance spectra simulation directly from molecular
  dynamics trajectories of a liquid crystal with a doped paramagnetic spin
  probe, Phys. Rev. Lett. 102 (2009) 013005.
\newblock \href {http://dx.doi.org/10.1103/PhysRevLett.102.013005}
  {\path{doi:10.1103/PhysRevLett.102.013005}}.

\bibitem{Oganesyan2011}
V.~S. Oganesyan, A general approach for prediction of motional epr spectra from
  molecular dynamics {(MD)} simulations: application to spin labelled protein,
  Phys. Chem. Chem. Phys. 13 (2011) 4724--4737.
\newblock \href {http://dx.doi.org/10.1039/C0CP01068E}
  {\path{doi:10.1039/C0CP01068E}}.

\bibitem{Budil2006}
D.~E. Budil, K.~L. Sale, K.~A. Khairy, P.~G. Fajer, Calculating slow-motional
  electron paramagnetic resonance spectra from molecular dynamics using a
  diffusion operator approach, J. Phys. Chem. A 110 (2006) 3703"1¤713.
\newblock \href {http://dx.doi.org/10.1021/jp054738k}
  {\path{doi:10.1021/jp054738k}}.

\bibitem{Blume}
M.~Blume, Stochastic theory of line shape: Generalization of the kubo-anderson
  model, Phys. Rev. 174 (1968) 351--358.
\newblock \href {http://dx.doi.org/10.1103/PhysRev.174.351}
  {\path{doi:10.1103/PhysRev.174.351}}.

\bibitem{Lenk}
R.~Lenk, Brownian Motion and Spin Relaxation, Elsevier, 1977.

\bibitem{GordonMess}
R.~G. Gordon, T.~Messenger, Electron Spin Relaxation in Liquids, Plenum, 1972.

\bibitem{Ito}
K.~It\^{o}, H.~McKean, Diffusion Processes and Their Sample Paths, Springer,
  1965.

\bibitem{Brillinger}
D.~R. Brillinger, A particle migrating randomly on a sphere, J. Theor. Probab.
  10 (1997) 165106.
\newblock \href {http://dx.doi.org/10.1007/978-1-4614-1344-8_7}
  {\path{doi:10.1007/978-1-4614-1344-8_7}}.

\bibitem{Sezerquater}
D.~Sezer, J.~H. Freed, B.~Roux, Simulating electron spin resonance spectra of
  nitroxide spin labels from molecular dynamics and stochastic trajectories, J.
  Chem. Phys. 128 (2008) 165106.
\newblock \href {http://dx.doi.org/10.1063/1.2908075}
  {\path{doi:10.1063/1.2908075}}.

\bibitem{Landau}
D.~Landau, E.~M. Lifshitz, Quantum Mechanics: Non-Relativistic Theory, Pergamon
  Press, 1981.

\end{thebibliography}
\end{document}